\documentclass[conference,compsoc]{IEEEtran}
\usepackage{amsmath}
\usepackage{graphicx}
\usepackage{algorithm,algpseudocodex}
\usepackage{multirow}
\usepackage{url}

\usepackage[flushleft]{threeparttable}
\usepackage{adjustbox}
\renewcommand{\arraystretch}{1.5}
\ifCLASSOPTIONcompsoc
  \usepackage[nocompress]{cite}
\else
  \usepackage{cite}
\fi
\graphicspath{{figures/}}

\begin{document}
%
\title{Detecting Compromised IoT Devices Using Autoencoders with Sequential Hypothesis Testing}

\author{\IEEEauthorblockN{Md Mainuddin, Zhenhai Duan}
\IEEEauthorblockA{Department of Computer Science\\
Florida State University\\
Tallahassee, FL 32306\\
Email: \{mainuddi, duan\}@cs.fsu.edu}
\and
\IEEEauthorblockN{Yingfei Dong}
\IEEEauthorblockA{Department of Electrical Engineering\\
University of Hawaii\\
Honolulu, HI 96822 USA\\
Email: yingfei@hawaii.edu\\
}}


%


\maketitle

\begin{abstract}
IoT devices fundamentally lack built-in security mechanisms to protect themselves from security attacks. Existing works on improving IoT security mostly focus on detecting anomalous behaviors of IoT devices. However, these existing anomaly detection schemes may trigger an overwhelmingly large number of false alerts, rendering them unusable in detecting compromised IoT devices. In this paper we develop an effective and efficient framework, named CUMAD, to detect compromised IoT devices. Instead of directly relying on individual anomalous events, CUMAD aims to accumulate sufficient evidence in detecting compromised IoT devices, by integrating an autoencoder-based anomaly detection subsystem with a sequential probability ratio test (SPRT)-based sequential hypothesis testing subsystem. CUMAD can effectively reduce the number of false alerts in detecting compromised IoT devices, and moreover, it can detect compromised IoT devices quickly. Our evaluation studies based on the public-domain N-BaIoT dataset show that CUMAD can on average reduce the false positive rate from about $3.57\%$ using only the autoencoder-based anomaly detection scheme to about $0.5\%$; in addition, CUMAD can detect compromised IoT devices quickly, with less than $5$ observations on average.
\end{abstract}


%
\IEEEpeerreviewmaketitle

\section{Introduction} \label{sec:intro}
In recent years Internet of Things (IoT) devices have been increasingly integrated into our daily lives and our society, with notable example environments such as smart homes, healthcare, transportation, and power grid. On one hand, this rapid development helps to improve the quality and efficiency of our daily lives. On the other hand, this same development also poses potentially unprecedented security and privacy challenges on the Internet, given that most of these IoT devices are low-cost systems with limited computation, memory, and energy resources. These devices often lack proper built-in security mechanisms to protect themselves and are vulnerable to various security attacks.

Many security attacks targeting or based on IoT devices have been reported in the past~\cite{pa2015iotpot}. In response to the growing problems of IoT security, government agencies such as US NIST have developed many recommendations that manufacturers should adopt to mitigate the security risks associated with IoT devices~\cite{Foundational2020Fagan}. In addition, many research efforts have been carried out to improve IoT security, including both proactive approaches to enhancing security mechanisms of IoT devices and more reactive solutions to monitor IoT device behaviors to detect rogue or infected IoT devices~\cite{al2020survey}.

Although some of the recommendations, for example, avoiding default common credentials, are relatively easy to be incorporated into IoT device manufacturing and certainly help mitigate IoT security risks, IoT devices are still fundamentally vulnerable to security attacks. As low-cost systems, IoT devices are inherently constrained in resources to support advanced security mechanisms. In addition, from the perspectives of both manufacturers and users, there are often conflicting objectives of IoT device usability and security, which often discourage the adoption of advanced security mechanisms in IoT devices.

Given these constraints of deploying advanced security mechanisms on IoT devices, network-based solutions have attracted a great amount of research efforts in recent years~\cite{al2020survey}. In particular, many machine learning (ML) based methods have been developed in detecting anomalous network behaviors of IoT devices~\cite{al2020survey}. (In this paper we use the term ML to refer to both traditional machine learning algorithms such as SVM and deep learning (DL) algorithms such as RNN.) However, most existing solutions only targeted the problem of anomaly detection in IoT devices~\cite{cook2019anomaly}, instead of detecting compromised IoT devices. Although detecting individual anomalies is of critical importance in certain application domains~\cite{chandola2009anomaly}, we note that these solutions may not be directly translated into the detection of compromised IoT devices. Given the large amount of network traffic, even a small false positive rate of an anomaly detection method can often translate into a large number of false alerts, rendering the detection method unusable in detecting compromised IoT devices in the real-world deployment. 

In this paper we develop an effective and efficient framework to detect compromised IoT devices, named CUMAD (cumulative anomaly detection). In essence, CUMAD integrates an autoencoder-based anomaly detection subsystem with a sequential probability ratio test (SPRT)-based sequential hypothesis testing subsystem~\cite{Goodfellow-et-al-2016,Wald1947:Sequential}. In CUMAD, the normal behavior of each IoT device is learnt and modeled by an autoencoder. During the training of an autoencoder model, it learns a latent space representation of the training data. More importantly, due to the nature of autoencoder, it excels at reconstructing inputs that are similar to the data used in training the model, but performs poorly when the new data is very different from the training data, manifested as large reconstruction errors. Although autoencoder has been mainly used in dimensionality reduction and feature learning in the past, in recent years it has also attracted a great amount of interests in anomaly detection in many different application domains. 

Instead of focusing on individual anomalous events detected by autoencoder, CUMAD aims to accumulate sufficient evidence to detect if an IoT device has been compromised. In CUMAD, the output of the autoencoder-based anomaly detection subsystem is fed into an SPRT-based sequential hypothesis testing subsystem. Unlike traditional probability ratio test methods that require a pre-defined fixed number of observations to reach a decision, SPRT works in an online manner and updates as observations arrive sequentially. SPRT reaches a conclusion whenever sufficient evidence has been observed. Therefore, SPRT can make a decision quickly (and consequently, CUMAD can detect compromised IoT devices quickly). 

In this paper we develop the CUMAD framework, and we also evaluate the performance of CUMAD using a public-domain IoT dataset N-BaIoT~\cite{meidan2018n}, which contains both benign and (Mirai and Bashlite) attack traffic of IoT devices. Our evaluation studies show that CUMAD can greatly improve the performance in detecting IoT devices in terms of false positive rates, for example, compared to the simple autoencoder-based anomaly detection system, CUMAD on average reduces the false positive rate from about $3.57\%$ to $0.5\%$, representing about $7$ times performance improvement in terms of false positive rate of the systems. In addition, CUMAD can detect a compromised IoT device quickly, with less than $5$ sequential observations on average. We note that although both autoencoder and SPRT have been proposed in developing anomaly detection systems before, to our knowledge, we are the first to integrate the two techniques to detect compromised IoT devices, instead of being used separately for anomaly detection. In addition, we are the first to introduce the notion of \emph{cumulative anomaly} in detecting compromised IoT devices (see Section~\ref{sec:related} for more details).

The remainder of the paper is organized as follows. In Section~\ref{sec:related} we discuss related work. We present the background on autoencoder and SPRT in Section~\ref{sec:back}. We describe the design of CUMAD in Section~\ref{sec:cumad}, and evaluate its performance in Section~\ref{sec:eval}. We conclude the paper in Section~\ref{sec:conc}.

\section{Related Work} \label{sec:related}
The problem of anomaly detection has been studied in many different application domains and many techniques have been proposed, based on statistical inference, data mining, signal processing, and recently machine learning, among others. We note that in the literature of anomaly detection, anomalies have been classified into three categories: point anomaly, contextual anomaly, and collective anomaly~\cite{chandola2009anomaly}. However, they are all concerned with the detection of individual anomalous events, which are different from the \emph{cumulative anomaly} we consider in this paper. In cumulative anomaly we are more concerned with the cause of anomalous events (for example, compromised IoT device), instead of individual anomalous events. As a consequence, we need to accumulate sufficient evidence (individual anomalous events) to reach a conclusion (for example, if an IoT device is compromised) in cumulative anomaly detection.

Given the importance of improving IoT security, many security attack detection techniques have been proposed, including various ML-based solutions~\cite{al2020survey,pang2021deep}. However, some of them required the training data of both benign and attack traffic. They cannot detect new security attacks. Others developed anomaly detection based schemes to detect anomalous traffic originated from IoT devices. However, as we have discussed in Section~\ref{sec:intro}, they often trigger a large number of false alerts, rendering them unusable in detecting compromised IoT devices in the real-world deployment.

In~\cite{gelenbe2022traffic}, Gelenbe and Nakip developed an online scheme CDIS to detect compromised IoT devices based on auto-associative learning. However, the design of CDIS was tailored to Mirai botnet, and may not be effective to detect other types of compromised IoT devices. In addition, CDIS still only targeted individual anomalous events, instead of cumulative anomaly detection as we perform in this paper.
The authors of~\cite{nguyen2019diot} developed a federated self-learning based scheme D{\"I}oT to detect compromised IoT devices, where local security gateways communicate with remote IoT Security Service to build a more comprehensive normal traffic model of IoT devices. In order to further reduce the false alerts generated by the aggregated anomaly detection model, a window-based scheme was adopted, where anomaly alarm was triggered only if the fraction of anomalous packets was greater than a pre-defined threshold value. 
In~\cite{meidan2018n}, Meidan \textit{et al.} presented an autoencoder-based anomaly detection system N-BaIoT to detect compromised IoT devices. N-BaIoT also tried to reduce the number of false alerts triggered by the pure anomaly detection system using a window-based scheme with a majority vote to reach a decision. 

%

\section{Background on Autoencoder and SPRT} \label{sec:back}
In this section we provide the necessary background on autoencoder and sequential probability ratio test (SPRT) for understanding the development of the proposed CUMAD framework. We refer interested readers to~\cite{Goodfellow-et-al-2016} and~\cite{Wald1947:Sequential}, respectively, for the detailed treatment on these two topics.

\begin{figure}[th]
\centering
               \includegraphics[scale=0.5]{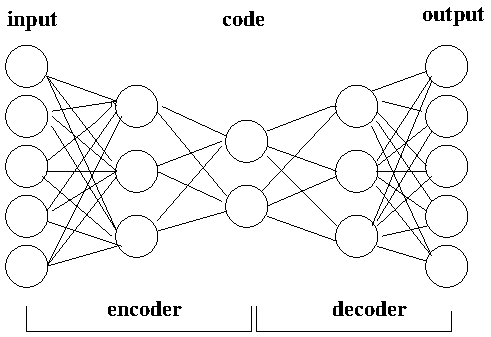}
                \caption{Illustration of Autoencoder.}
                \label{fig:autoencoder}
\end{figure}
\subsection{Autoencoder}
Autoencoder is an unsupervised neutral network that aims to reconstruct the input at the output. Figure~\ref{fig:autoencoder} illustrates a simple standard (undercomplete) autoencoder. An autoencoder can be considered as consisting of two components: an encoder $f$ and an decoder $g$. Given input data $\mathbf{x}$, the encoder function $f$ maps $\mathbf{x}$ to a latent-space representation, or code $\mathbf{h}$, that is $\mathbf{h} = f(\mathbf{x})$. Using the corresponding code $\mathbf{h}$ as the input, the decoder function $g$ tries to reconstruct the original input $\mathbf{x}$ at its output $\mathbf{x'}$, that is, $\mathbf{x'} = g(\mathbf{h})$. Combining both the encoder function and decoder function together, we have $\mathbf{x'} = g(f(\mathbf{x}))$. Let $L(\mathbf{x}, \mathbf{x'})$ be the reconstruction error, that is, the difference between $\mathbf{x}$ and $\mathbf{x'}$. The autoenceder aims to minimize $L(\mathbf{x}, \mathbf{x'})$. We note that there are different definitions of $L(\mathbf{x}, \mathbf{x'})$ and one of the most common definitions is the \textit{mean squared errors} (MSE). We note that in the example autoencoder of Figure~\ref{fig:autoencoder}, both the encoder and decoder have only one hidden layer. This is only for illustration purpose. In reality they can have many hidden layers, depending on the specific application requirement.

Autoencoders have been traditionally used in applications of dimensionality reduction and feature learning, by focusing on the compressed code of an autoencoder, which holds the latent-space representation of the original data. On the other hand, autoencoders also possess a few desired properties, making them an attractive candidate for anomaly detection. For example, an autoencoder is able to extract the salient features of the original data to remove dependency in the original data. More importantly, an autoencoder can only learn the properties or distributions of the data that it has seen during the training stage, that is, the data points in the training dataset. It excels at reconstructing data that are similar to the training data, but performs poorly on data that are very different from the training data, in terms of the reconstruction error $L(\mathbf{x}, \mathbf{x'})$.

This is an appealing property of autoencoders in the application of anomaly detection. For example, in the context of detecting compromised IoT devices, we can establish the normal behavioral model of an IoT device using an autoencoder by training it with benign network traffic before the device has been compromised. We can continue monitoring the IoT device by passing the corresponding network traffic of the device into the trained model. If the reconstruction error is no greater than a pre-specified threshold, we consider the corresponding network traffic to be benign. When the reconstruction error is greater than the threshold, we claim that the network traffic is anomalous.

\subsection{Sequential Probability Ratio Test}
Sequential probability ratio test (SPRT) is a simple yet powerful statistical tool that has found applications in many different domains, in particular, fault detection or quality control~\cite{Wald1947:Sequential}. SPRT is a variant of the traditional probability ratio test for testing under what distribution (or with what distribution parameters), it is more likely to have the observed sequence of samples. Unlike traditional probability ratio test that requires a pre-defined fixed number of samples to carry out the test, SPRT works in an online fashion; it updates the corresponding statistical measure as samples arrive sequentially, and can conclude when sufficient samples have arrived to reach a decision. In its simplest form, SPRT is a statistical method to test a simple null hypothesis against a simple alternative hypothesis. In the following we will more formally describe the operation of SPRT.

Let $\mathbf{y}$ denote a Bernoulli random variable with an unknown parameter $\theta$, and let $\mathbf{y_i}$, for $i = 1, 2, \ldots$ denote the corresponding successive observations of $\mathbf{y}$. SPRT can be used to test a simple null hypothesis $H_0$ that $\theta = \theta_0$ against a simple alternative hypothesis $H_1$ that $\theta = \theta_1$, more specifically,
\begin{align*}
Pr(\mathbf{y_i} = 1|H_0) &= 1 - Pr(\mathbf{y_i} = 0|H_0) = \theta_0 \\
Pr(\mathbf{y_i} = 1|H_1) &= 1 - Pr(\mathbf{y_i} = 0|H_1) = \theta_1.    
\end{align*}

As observations $\mathbf{y_i}$ for $i = 1, 2, \dots, n$ arrive one by one, SPRT maintains a probability ratio measure, namely, 
\[
\frac{Pr(\mathbf{y_1, y_2,\ldots,y_n}|H_1)}{Pr(\mathbf{y_1, y_2,\ldots,y_n}|H_0)}
\]
to test under which hypothesis ($H_1$ or $H_0$) it is more likely to observe the sequence of samples $\mathbf{y_i}$, for $i = 1, 2, \ldots, n$.
In order to simplify exposition and computation, in the following we will compute the logarithm of the probability ratio and denote it as $\Lambda_n$:

\begin{equation*}
\Lambda_n = ln\frac{Pr(\mathbf{y_1, y_2,\ldots,y_n}|H_1)}{Pr(\mathbf{y_1, y_2,\ldots,y_n}|H_0)}.
\end{equation*}

Assume that $\mathbf{y_i}$'s are independent (and identically distributed), we have
\begin{equation}\label{eq:lambda}
\Lambda_n = ln\frac{\prod_1^nPr(\mathbf{y_i}|H_1)}{\prod_1^nPr(\mathbf{y_i}|H_0)} =
\sum_{i=1}^nln\frac{Pr(\mathbf{y_i}|H_1)}{Pr(\mathbf{y_i}|H_0)} = \sum_{i=1}^nz_i 
\end{equation}

where $z_i = ln\frac{Pr(\mathbf{y_i}|H_1)}{Pr(\mathbf{y_i}|H_0)}$, which can be considered as
the step in the random walk represented by $\Lambda$. When the
observation is one ($\mathbf{y_i} = 1$), the constant $ln\frac{\theta_1}{\theta_0}$
is added to the preceding value of $\Lambda$. When the observation is
zero ($\mathbf{y_i} = 0$), the constant $ln\frac{1-\theta_1}{1-\theta_0}$ is added.

Now we are ready to describe the operation of SPRT for testing $H_0$ against $H_1$. As a sample $\mathbf{y_i}$ arrives, depending on the value of the sample, the statistical measure $\Lambda_n$ is updated according to Eq.~(\ref{eq:lambda}). The measure $\Lambda_n$ is then compared against two user-defined thresholds $A$ and $B$ ($A < B$) to check if SPRT can reach a decision or more samples are needed:
\begin{align} \label{eq:sprt_decision}
\Lambda_n \leq A &\Longrightarrow \mbox{ accept $H_0$ and terminate test}, \nonumber\\
\Lambda_n \geq B &\Longrightarrow \mbox{ accept $H_1$ and terminate test}, \\
A < \Lambda_n < B &\Longrightarrow \mbox{ take an additional observation}. \nonumber
\end{align}

The two thresholds $A$ and $B$ in the operation of SPRT can be approximated using the user-desired false positive rate $\alpha$ and false negative rate $\beta$~\cite{Wald1947:Sequential}:
\begin{equation}
A \approx ln\frac{\beta}{1-\alpha}, \mbox {   } B \approx
ln\frac{1-\beta}{\alpha}. \label{eq:boundary}
\end{equation}

As a simple and powerful statistical tool, SPRT possesses a few critical and desired properties that lead to the wide-spread application of the technique in many different domains. First, the false positive and false negative rates of SPRT can be specified by user-desired error rates, which in turn control the thresholds of the model. Second, it has been proved that, among all sequential and non-sequential probability ratio testing algorithms, SPRT minimizes the expected number of observations to reach a decision with no greater errors. Put in another way, on average SPRT can reach a conclusion quickly compared to other probability ratio testing algorithms.

In the context of anomaly detection, we can consider $\mathbf{y_i} = 1$ as an anomalous sample, and $\mathbf{y_i} = 0$ as a normal sample. Instead of simply relying on a single sample to conclude the nature of the sample's origin (being normal or abnormal), SPRT continues observing additional samples $\mathbf{y_{i+1}, y_{i + 2}, \ldots}$ until sufficient evidence is collected to reach a conclusion. In essence, SPRT infers the distribution or the parameters of the distribution of a random process, based on sequential observations of samples drawn from the random process. This matches our objectives of CUMAD well, where we would like to accumulate sufficient evidence to detect if an IoT device is compromised, instead of an individual anomalous sample. 


\section{Design of CUMAD} \label{sec:cumad}
In this section we will first discuss the considered network model, where CUMAD will be deployed, and then we will present the design of the CUMAD framework.
\begin{figure}[th]
\centering
               \includegraphics[scale=0.55]{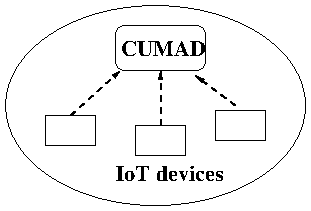}
                \caption{Conceptual network model.}
                \label{fig:network}
\end{figure}

\subsection{Network Model}
Figure~\ref{fig:network} illustrates the conceptual network model, where CUMAD is deployed. As shown in the figure, in order for CUMAD to carry out its task to detect compromised IoT devices in a network, CUMAD needs to have access to the network traffic associated with the IoT devices in the network. Depending on the deployment scenarios of CUMAD in the network and the corresponding network architecture, there can be a few different ways for CUMAD to obtain the corresponding network traffic of IoT devices. In essence, CUMAD as a network-based solution can be deployed in a similar way as network-based intrusion detection systems.

In the current design of CUMAD, (statistical) features from raw network traffic will be extracted and fed to CUMAD for detecting compromised IoT devices. Each input data point fed to CUMAD comprises these extracted features, and can be summarized at different levels of granularity of network traffic, such as packets, flows, and time windows. These features will capture the network behavioral characteristics of the corresponding IoT devices. In Section~\ref{sec:eval} we will discuss the network traffic features contained in the public-domain N-BaIoT dataset when we perform evaluation studies on CUMAD~\cite{meidan2018n}.

\begin{figure*}[th]
\centering
               \includegraphics[scale=0.55]{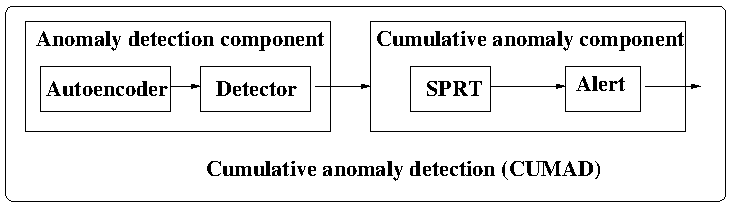}
                \caption{Illustration of CUMAD architecture.}
                \label{fig:cumad}
\end{figure*}
\subsection{CUMAD: Cumulative Anomaly Detection}
Figure~\ref{fig:cumad} illustrates the high-level architecture of the CUMAD framework. CUMAD consists of two main components: an anomaly detection component (ADC) and a cumulative anomaly component (CAC). Assuming the model has been properly trained (will be discussed shortly), given an input data point with the corresponding features, the main responsibility of ADC is to classify an input data point as either normal or anomalous. After the classification of the input data point, the result is passed to the second component (CAC), which will maintain a cumulative view of the network traffic behavior of the corresponding IoT device, by sequentially merging the individual classification results into the view. When sufficient evidence on an IoT device has been collected to indicate that it has been compromised, an alert will be generated. In the following we will describe each component in details, both in model training and deployment to detect compromised IoT devices. 

We note that different types of IoT devices perform drastically different functionalities, and in addition, we would like to detect which IoT device is compromised, we need to develop a separate CUMAD model for each IoT device and monitor their network traffic behaviors separately using their corresponding CUMAD models. Therefore, the following discussions are for one IoT device. We note that, although there are vastly diverse types of IoT devices on the Internet, autoencoder is a powerful neural network that is capable of learning different models. Therefore, we are able to build diverse autoencoder models, one for each IoT device, despite their vastly different network traffic behaviors of these IoT devices. In addition, IoT devices also provide us with unique opportunities in establishing the models of normal behaviors, compared to traditional computer systems. In particular, each IoT device only performs a few well-defined simple functionalities in an autonomous or semi-autonomous fashion, with very limited user interactions after the initial device configuration and setup. This makes it simpler to establish a model of normal behaviors in carrying out anomaly detection.

\subsubsection{Model Training and Setup}
Before CUMAD can be used to monitor network traffic to detect compromised IoT devices, we need to train a CUMAD system for each IoT device so that it can learn the normal model of the device. During the training stage of a deployed CUMAD system, it is critical that we should only feed normal (benign) network traffic of the device to the system. This can be done, for example, when an IoT device is first deployed in the network. In order to minimize false positives during the detection stage, it is also important that CUMAD has a reasonably complete view of all the normal network traffic behavior of the device. 

The training of the autoencoder module in ADC (see Figure~\ref{fig:cumad}) follows the basic standard steps in ML training~\cite{chollet2021deep}. Given a set $\mathbf{D}$ of normal input data points to the autoencoder of ADC , we partition it into two subsets: the training set $\mathbf{D_t}$ and the validation set $\mathbf{D_v}$. We use $\mathbf{D_t}$ to train the model and use $\mathbf{D_v}$ to obtain the optimal value of the hyperparameters including learning rate and epochs to optimize the performance of the final model in terms of reconstruction errors. After the optimal values of these hyperparameters are obtained, we perform the final training of the model using $\mathbf{D}$. After the training is complete, the autoencoder obtains the latent-space representation of the normal network traffic behavior of the corresponding IoT device, which can capture the salient properties of the device's network traffic.

As discussed above, the premise of using an autoencoder as an anomaly detection mechanism is that, although it can effectively reconstruct data points that are similar to the data points that it has seen previously during the training stage, it in general performs poorly to reconstruct data points that substantially differ from the training data. This is manifested in large reconstruction errors. Therefore, we will use the reconstruction error as the anomaly score, and when the anomaly score is greater than the pre-defined threshold, we classify the corresponding input data point as an anomalous sample. 

Clearly, the choice of the anomaly score threshold $T_{as}$ will greatly affect the performance of the resulting CUMAD system. In the current design, we choose $T_{as}$ in the following manner. Let $\mu_D$ and $\sigma_D$ denote the average and standard deviation of the mean squared errors for data points in the dataset $\mathbf{D}$, respectively, then $T_{as} = \mu_D + \sigma_D$. This threshold value is maintained by the Detector module in ADC (see Figure~\ref{fig:cumad}).

The training of the autoencoder and the choice of $T_{as}$ also provide us with some hints on the parameters of SPRT in CAC. In particular, based on the value of $T_{as}$ we can compute the proportion of input data points in $\mathbf{D}$ that will be (mistakenly) classified as being anomalous, and we can treat this proportion as the value for $\theta_0$, that is, the probability that we will have an anomalous sample when the device is not compromised. In addition to the parameter $\theta_0$, we also need to set up a few other parameters for SPRT to work, including $\theta_1$, $\alpha$, and $\beta$. $\theta_1$ is the probability that an input data point is anomalous when the hypothesis $H_1$ is true, that is, when the IoT device is compromised. During the time when a compromised IoT device is used to launch a security attack, this probability should be much higher than $\theta_0$. Although in general we cannot obtain the precise value of $\theta_1$, a value sufficiently greater than $\theta_0$ should work. We note that imprecise values of $\theta_1$ (and $\theta_0$) may result in a larger number of samples for SPRT to reach a decision. 

The parameters $\alpha$ and $\beta$ are the user-desired false positive rate and false negative rate, respectively. They normally have small values for all practical applications, for example, in the range $0.01$ to $0.05$. The initial value of $\Lambda_n$ in Eq.~(\ref{eq:lambda}) is set to $0$ during the setup stage of the system. The functionality of the Alert module is to generate proper alert to inform system administrators of the detection of a compromised IoT device. Other actions can also be taken based on the local security policies, for example, informing proper agents to isolate the compromised IoT device.

\begin{algorithm}[th]
\caption{CUMAD procedures in detection.} \label{alg:cumad-detection}
\begin{algorithmic}[1]
\State Let $\mathbf{x}$ be an input data point
\Procedure{Autoencoder}{$\mathbf{x}$}
\State $\mathbf{x'} \leftarrow g(f(\mathbf{x}))$
\Comment{Reconstruct $\mathbf{x}$ in autoencoder}
\State $s \leftarrow L(\mathbf{x}, \mathbf{x'})$
\Comment{Compute anomaly score}
\EndProcedure

\Procedure{Detector}{$s$}
\If{$s > T_{as}$}
\State return 1
\Comment{Anomalous input data point}
\Else
\State return 0
\Comment{Normal input data point}
\EndIf
\EndProcedure

\Procedure{SPRT}{$o$}
\State $o$: output of Detector
\LComment{Update $\Lambda_n$ based on Eq.~(\ref{eq:lambda}) }
  \If {($o == 1$)}
  \LComment{anomalous}
  \State $\Lambda_n += ln\frac{\theta_1}{\theta_0}$
   \Else
  \LComment{normal}
  \State $\Lambda_n += ln\frac{1-\theta_1}{1-\theta_0}$
  \EndIf
  \LComment{If decision can be made based on Eq.~(\ref{eq:sprt_decision})}
  \If {($\Lambda_n \geq B$)}
  \State Device is compromised. 
  \State Test terminates for the IoT device
  \State Calling Alert module to generate alert
  \ElsIf {($\Lambda_n \leq A$)}
  \State Device is normal. Test is reset for the device
  \State $\Lambda_n \leftarrow 0$
  \State Test continues with new observations
  \Else
  \State Test continues with an additional observation
  \EndIf
\EndProcedure

\end{algorithmic}
\end{algorithm}

\subsubsection{Detection}
After the model has been trained and the required parameters have been set for the CUMAD system, it can be used to monitor network traffic to detect if the corresponding IoT device has been compromised. In the following we describe the basic steps of a CUMAD system in carrying out the detection task (see Algorithm~\ref{alg:cumad-detection}).

The input data point $\mathbf{x}$ is fed into the trained autoencoder model of CUMAD (lines $2$ to $4$ in Algorithm~\ref{alg:cumad-detection}). The autoencoder will try to reconstruct the input and create the output $\mathbf{x'}$. The anomaly score $s$ is then computed (it is the mean squared error in the current design). The anomaly score is then passed to the Detector module of ADC, which maintains an anomaly score threshold to distinguish between normal and anomalous data points (lines $5$ to $9$ in the algorithm). If the anomaly score is greater than the threshold $T_{as}$, an anomalous data point is identified, and an output value $1$ is generated in the Detector to indicate the detection of the anomalous data point. Otherwise, an output value $0$ is generated to indicate a normal data point.

The output of the Detector module is then passed to the SPRT module to determine if sufficient evidence has been accumulated to make a decision regarding the nature of the IoT device (compromised or normal; line $10$ of the algorithm). SPRT updates the probability ratio measure $\Lambda_n$ according to Eq.~(\ref{eq:lambda}), as the $0$ (normal data point) and $1$ (anomalous data point) output sequence of the Detector module arrives (lines $13$ to $18$). After the value of $\Lambda_n$ is updated for each input data point, SPRT compares the value of $\Lambda_n$ with the two boundaries $A$ and $B$ to determine if a decision can be made (lines $20$ to $29$). When the value of $\Lambda_n$ hits or crosses the upper bound $B$, SPRT will conclude that the alternative hypothesis $H_1$ is true, that is, the IoT device has been compromised. In this case, SPRT will inform the Alert module the detection of an compromised IoT device. Proper alert will be generated and corresponding system administrators will be informed. In addition, from this time on, it is not necessary for CUMAD to monitor the IoT device anymore, until proper actions have been taken to clean up or remove the device. 

When the value of $\Lambda_n$ is equal to or smaller than the lower bound $A$, SPRT reaches the conclusion that $H_0$ is true, that is, the IoT device is not compromised. From the viewpoint of detecting compromised IoT devices, this conclusion is less interesting in that we cannot terminate the monitoring of the device as we have done when a compromised IoT device is detected. A normal IoT device may become compromised at a later time. Therefore, in this case, we will reset the state of SPRT to restart the monitoring of the IoT device, in particular, we will reset the value of $\Lambda_n$ to zero. If a decision cannot be reached at this time (line $28$), SPRT will simply wait for additional input data points and repeat the same procedure.

\section{Evaluation Studies} \label{sec:eval}
In this section we perform evaluation studies to investigate the performance of CUMAD using the public-domain N-BaIoT dataset~\cite{meidan2018n}. In order to better understand the evaluation studies, we will first describe the dataset, in particular, the features of the data points contained in the dataset. We will also compare the performance of CUMAD with that of the N-BaIoT scheme (which is the name for both the dataset and the corresponding scheme on detecting compromised IoT devices)~\cite{meidan2018n}.

\subsection{Dataset, Features, and CUMAD System Setup}

\begin{table}[th]
\renewcommand{\arraystretch}{1.5}
\caption{N-BaIoT feature extraction aggregations.}
\label{table:anomaly-nbaiot-feature-extraction}
\centering
\begin{threeparttable}[htbp]
\begin{adjustbox}{width=\columnwidth,center}
\begin{tabular}{|l|l|l|l|l|}
  \hline
  Aggregation level & \# features & Pkt attributes and measures \\
  \hline
  Source IP & 3 & \# pkts, Mean/Variance of pkt sizes\tnote{1} \\
  \hline
  Source MAC-IP & 3 & \# pkts, Mean/Variance of pkt sizes\tnote{1} \\
  \hline
  \multirow{2}{*}{Channel} & 10 & \# pkts, Mean/Variance of pkt sizes\tnote{1}, Mean/Variance/Count of IATs, \\
  & & Magnitude/Radius/Covariance/Correlation Coefficient of pkt sizes\tnote{2} \\
  \hline
  
  \multirow{2}{*}{Socket} & 7 & \# pkts, Mean/Variance of pkt sizes\tnote{1},  \\
  & & Magnitude/Radius/Covariance/Correlation Coefficient of pkt sizes\tnote{2} \\
  \hline
\end{tabular}
\end{adjustbox}
\begin{tablenotes}
\footnotesize
\item[1]Outgoing packets only
\item[2]Both incoming and outgoing packets
\end{tablenotes}
\end{threeparttable}
\end{table}

\begin{table*}[th]
\centering
\renewcommand{\arraystretch}{1.5}
\caption{Performance Results}
\label{tab:performance-summary}
\resizebox{\textwidth}{!}{
\begin{tabular}{|l|l|l|l|l|l|l|l|l|}
  \hline
   \multirow{2}{*}{Device} & \multicolumn{4}{c|}{CUMAD} & \multicolumn{4}{c|}{N-BaIoT} \\
  \cline{2-9}
   &  Mean Size & Accuracy & Recall & F1 Score & Window Size & Accuracy & Recall & F1 Score \\
    \hline
     Danmini Doorbell & 5.19 & 0.979 & 1.000 & 0.979 & 82 & 0.995 & 1.000 & 0.995 \\
    \hline
    Ecobee Thermostat & 4.18 & 0.988 & 1.000 & 0.988 & 20 & 0.995 & 1.000 & 0.995 \\
    \hline
    Philips B120N10 Baby Monitor & 5.62 & 0.955 & 0.994 & 0.957 & 65 & 0.982 & 0.992 & 0.983 \\
    \hline
    Provision PT 737E Security Cam & 4.21 & 0.992 & 1.000 & 0.992 & 32 & 1.000 & 1.000 & 1.000 \\
    \hline
    Provision PT 838 Security Cam & 4.79 & 0.957 & 1.000 & 0.958 & 43 & 0.978 & 1.000 & 0.978 \\
    \hline
    SimpleHome XCS7 1002 WHT Security Cam & 4.08 & 0.995 & 0.999 & 0.995 & 23 & 0.998 & 0.998 & 0.998 \\
    \hline
    SimpleHome XCS7 1003 WHT Security Cam & 4.11 & 0.993 & 1.000 & 0.993 & 25 & 0.995 & 1.000 & 0.995 \\
    \hline
  \end{tabular}
  }
\end{table*}

N-BaIoT contains both benign and (Mirai and Bashlite) attack traffic of $9$ commercial IoT devices, including two doorbells (Danmini and Ennio), an Ecobee thermostat, three baby monitors (different models from Provision and Philips), two SimpleHome security cameras, and a Samsung webcam. Benign IoT device traffic was collected immediately after the corresponding IoT device was connected to the experimental testbed. Care was taken to ensure that various representative normal operations and behaviors of IoT devices were collected into the benign dataset.   

These IoT devices were later infected with Mirai and Bashlite malware in the controlled environment, and the attack traffic was also collected and added into the dataset. The dataset does not contain Mirai attack data for two of the devices (Ennio Doorbell and Samsung Webcam). We exclude these two IoT devices from our evaluation studies, therefore, we have $7$ IoT devices in our evaluation studies. In our evaluation studies, we use the benign traffic and the scan attack traffic from the remaining $7$ IoT devices. We equally partition the benign data of an IoT device into three datasets: the training dataset $\mathbf{D_t}$, the validation dataset $\mathbf{D_v}$, and the remaining $1/3$ of benign data is merged with the same number of attack data points to form the balanced test dataset $\mathbf{D_{tst}}$.

In the N-BaIoT dataset, each data point corresponds to an arrived packet, and contains $115$ statistical features, which together represent a behavioral snapshot that describes the context of the corresponding packet when it arrives at the data collection point. The snapshot contains the source and destination device information, the protocol information, among others. More specifically, the $115$ features were extracted in the following manner. For each arriving packet, a total of $23$ features were collected at different levels of aggregation (see Table~\ref{table:anomaly-nbaiot-feature-extraction}), including features aggregated at source IP address level, at source MAC and IP addresses level, at level of channel (source and destination IP addresses), and at socket level (source and destination IP addresses and port numbers). These $23$ features were extracted in a sliding window fashion, over $5$ time windows of $100$ms, $500$ms, $1.5$sec, $10$sec, and $1$min, respectively, generating a total of $115$ features for each data point.

We use the Keras sequential model as the foundation for our development of the autoencoder~\cite{chollet2021deep}. The model's input dimension is set to match the number of features in the dataset (that is, $115$). To ensure effective compression, we implement three hidden layers within the encoder. These layers progressively reduce the dimensions to $87$, $58$, $38$, and $29$, respectively, with the last one ($29$) being the dimension of the output layer of the encoder, that is, the dimension of the obtained code. Conversely, the decoder component mirrors the dimensions of the encoder layers in the reverse order, starting from $38$. By employing compression and decompression in the encoder and decoder layers, we effectively eliminate redundant information from the features of the input data points. To optimize training performance, we utilize the Adam optimizer, and the mean square error is used as the reconstruction error (objective function of the model). 

SPRT requires four user-defined parameters in order to compute the upper and lower bounds $A$ and $B$ (see Eq.~(\ref{eq:boundary})), as well as the step function for computing $\Lambda_n$ following each observation (see Eq.~(\ref{eq:lambda})). The desired values for both the false positive rate and the false negative rate (represented by $\alpha$ and $\beta$, respectively) are typically very small. In this study we set both $\alpha$ and $\beta$ to $0.01$. Ideally, the parameter $\theta$ indicates the true probability of an observation being classified as an anomaly, from either a benign or compromised IoT device. We determine the values for $\theta_0$ and $\theta_1$ through our preliminary studies, and set them to $0.2$ and $0.8$, respectively.

\subsection{Performance Results}
Table~\ref{tab:performance-summary} shows the performance of CUMAD in detecting IoT devices, in terms of accuracy, recall, and F1 score~\cite{chollet2021deep}. From the table we can see that CUMAD achieves superior performance in all three metrics. For example, for $5$ of the IoT devices, CUMAD is able to detect all the compromised cases (see the column of Recall). CUMAD is also able to detect vast majority of the compromised cases for the remaining two of the IoT devices, with recall scores of $0.999$ and $0.994$. Considering both detection precision of attack and benign traffic, we can see that CUMAD also performs very well, with an accuracy score ranging from $0.955$ to $0.995$ for all $7$ IoT devices. The F1 scores, which is a weighted average of the precision and recall scores of a model, also confirm that CUMAD performs well in detecting compromised cases.

\begin{figure}[th]
    \centering
    \includegraphics[width=0.8\columnwidth]{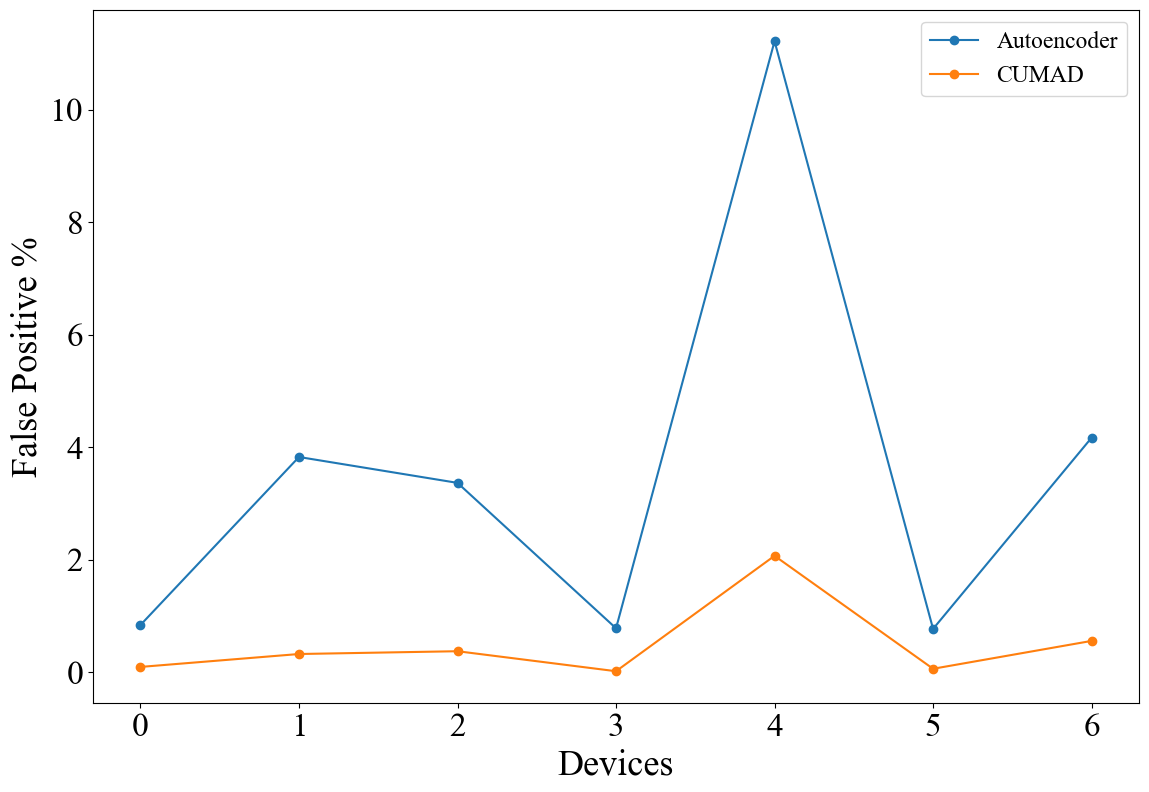}
    \caption{False positive rates.}
    \label{fig:false-positive}
\end{figure}

Figure~\ref{fig:false-positive} shows the false positive rates of an autoencoder-based anomaly detection scheme and CUMAD. As shown in the figure, the false positive rates of the autoencoder-based anomaly detection scheme for the $7$ IoT devices range from $0.77\%$ to $11.22\%$, while the false positive rates of CUMAD range from $0.014\%$ to $2.067\%$. On average the autoencoder-based anomaly detection scheme has about $3.57\%$ false positive rate, while the false positive rate of CUMAD is about $0.5\%$, which represents about $7$ times performance improvement in terms of false positive rate for CUMAD over the autoencoder-based anomaly detection scheme.

\begin{figure}[th]
    \centering
    \includegraphics[width=0.8\columnwidth]{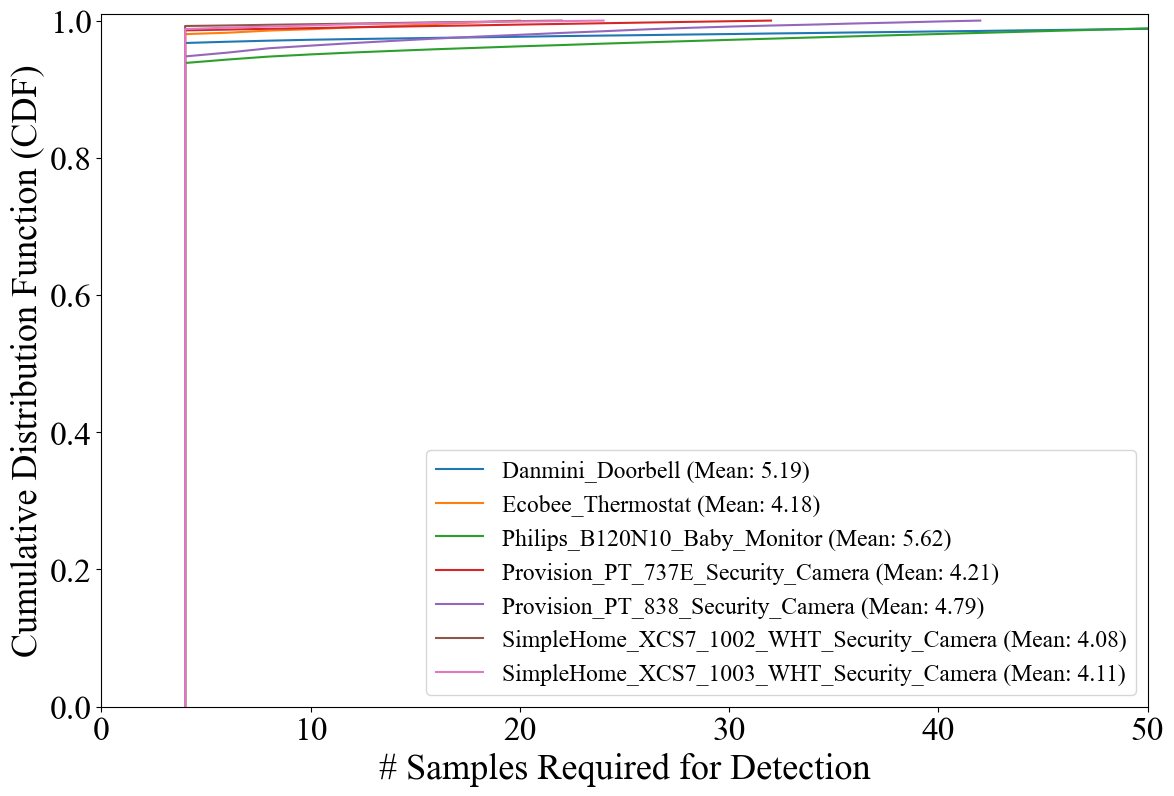}
    \caption{\# of observations for detection in CUMAD.}
    \label{fig:required-window-sizes}
\end{figure}

For performance comparison, we also include in the table the performance results of the N-BaIoT scheme, with the same evaluation studies setup. We can see from the table that CUMAD and N-BaIoT performs comparably in terms of all three performance metrics. However, N-BaIoT works on a fixed window size. Table~\ref{tab:performance-summary} shows that N-BaIoT requires a relatively large window size, ranging from $20$ to $82$ (column with name {\em Window Size}). In contrast, CUMAD works in an online fashion and does not requires such a fixed window size. Table~\ref{tab:performance-summary} shows the average number of observations required for CUMAD to reach a detection (column with name {\em Mean Size}); we can see from the table that it takes on average less than $5$ observations for CUMAD to make a detection of a compromised case, much quicker than N-BaIoT. In order to have a better understanding of the number of observations for CUMAD to make a detection of a compromised case, Figure~\ref{fig:required-window-sizes} shows the cumulative distribution function (CDF) of required observations for CUMAD to make a detection for all the $7$ IoT devices. We can see from the figure that the vast majority of detection requires less than $10$ observations for all $7$ IoT devices.

In summary, compared to simple anomaly detection schemes such as the ones only based on autoencoders, CUMAD can greatly reduce the false positive rates, making CUMAD much more attractive than simple anomaly detection schemes in the real-world deployment. Compared with window-based schemes such as N-BaIoT, CUMAD requires much less observations to reach a detection, and thus can detect compromised IoT devices much quicker.

\section{Conclusions} \label{sec:conc}
In this paper we have developed CUMAD, a cumulative anomaly detection framework for detecting compromised IoT devices. CUMAD employs an unsupervised neural network autoencoder to classify whether an individual input data point is anomalous or normal. CUMAD also incorporates a statistical tool sequential probability ratio test (SPRT) to accumulate sufficient evidence to detect if an IoT device is compromised, instead of directly relying on individual anomalous input data points. CUMAD can greatly improve the performance in detecting compromised IoT devices in terms of false positive rate compared to the methods only relying on individual anomalous input data points. In addition, as a sequential method, CUMAD can quickly detect compromised IoT devices. Evaluation studies based on public-domain IoT dataset N-BaIoT confirmed the superior performance of CUMAD.
\bibliographystyle{unsrt}
\bibliography{net, iot}
%



\end{document}